# Evaluating Performance and Gameplay of Virtual Reality Sickness Techniques in a First-Person Shooter Game

Diego Monteiro*,
Department of Computing
Xi'an Jiaotong-Liverpool University
Suzhou, China
D.Monteiro@ xjtlu.edu.cn;

DMT Lab
Birmingham City University
Birmingham, UK
Diego.VilelaMonteiro@bcu.ac.uk;
* Corresponding author

Hao Chen , Hai-Ning Liang*,
Department of Computing
Xi'an Jiaotong-Liverpool University
Suzhou, China
Hao.Chen17@student.xjtlu.edu.cn
HaiNing.Liang@ xjtlu.edu.cn;
* Corresponding author

Huawei Tu, Henry Duh
Computer Science and Information Technology
LaTrobe University
Melbourne, Australia
{H.Tu, B.Duh}@latrobe.edu.au

*Abstract* - **In virtual reality (VR) games, playability and immersion levels are important because they affect gameplay, enjoyment, and performance. However, they can be adversely affected by VR sickness (VRS) symptoms. VRS can be minimized by manipulating users' perception of the virtual environment via the head-mounted display (HMD). One extreme example is the Teleport mitigation technique, which lets users navigate discretely, skipping sections of the virtual space. Other techniques are less extreme but still rely on controlling what and how much users see via the HMD. This research examines the effect on players' performance and gameplay of these mitigation techniques in fast-paced VR games. Our focus is on two types of visual reduction techniques. This study aims to identify specifically the trade-offs these techniques have in a first-person shooter game regarding immersion, performance, and VRS. The main contributions in this paper are (1) a deeper understanding of one of the most popular techniques (Teleport) when it comes to gameplay; (2) the replication and validation of a novel VRS mitigation technique based on visual reduction; and (3) a comparison of their effect on players' performance and gameplay.**

## I. INTRODUCTION

In virtual reality (VR) games, playability and immersion are two primary factors because they affect gameplay, enjoyment, and performance. However, these two factors can be adversely affected by VR sickness (VRS) symptoms. Despite recent advances in VR, VRS remains a significant challenge to overcome, especially in games, because it is often difficult to reduce it without compromising in-game performance, gameplay, and experience[1].

One technique that has been shown to mitigate some VRS symptoms is Teleport [2]. Teleport is a well-established technique that effectively allows users to skip discretely whole sections of the virtual environment and has presented excellent results in reducing nausea and oculomotor symptoms. On the other hand, its trade-offs often involve performance loss, which can be acceptable in some contexts but might not be desirable in games. This issue led us to our first research question: (*RQ1*) *What are the trade-offs in immersion, VRS, and player performance when using Teleport in games, especially fast-paced ones like first-person shooters (FPS)?*

Recently researchers have proposed other techniques that can help reduce VRS. These techniques manipulate what users perceive in the head-mounted display (HMD) and how they see elements in the 3D virtual environment [1], [3]. Unavoidably, they also have their trade-offs. For example, some techniques use field-of-view (FoV) reduction to mitigate VRS ([4], [5]). More recently, others have attempted to provide a copy of the 3D view and display it as a 2D view or 2D frame (2DF for short) [6]. That is, it reduces the amount of visual detail that users see. This technique has been shown to produce the same level of immersive experience in a VR FPS game but with reduced levels of VRS and without affecting players' performance and gameplay in the fast-paced game.

Teleport has not yet been compared to these newer techniques, like 2DF. The positive results of 2DF led us to explore our next two research questions: (*RQ2*) *Is Teleport a more effective VRS mitigation technique than a technique like 2DF, which reduces aspects of the environment but still allows continuous movement?* (*RQ3*) *Can gamers using Teleport perform better than using a technique like 2DF?*

This research aims to answer these three questions to help understand what types of techniques are more useful in a fast-paced environment, like an FPS game, in VR. To study the effects of 2DF as an alternative to Teleport in FPS games, we replicated the technique used in [6] and used it in a similar VR environment. Our results show that Teleport offered a worse trade-off than 2DF in player performance and provided similar levels of VRS. Our results suggest that a technique like 2DF could be a valuable alternative to Teleport in FPS games and other similar VR applications.

The main contributions of this paper are: (1) a deeper understanding of one of the most popular techniques (Teleport [2]) in games; (2) the replication and validation of a new VRS mitigation technique that reduces visual detail [6]; and (3) a comparison of their effect on player performance and gameplay.

## II. BACKGROUND

VRS is very common and has been studied for quite some time [7]. Despite this, its origins and sources are still not fully understood. Two theories that have broad support are the Rest Frame Hypothesis [8], [9], and the Sensory Conflict Theory [10]. The Rest Frame Hypothesis posits that the human brain assumes that some stationary environmental objects can serve as reference points to view the world. Accordingly, using fixed gazing points or other visual objects [11], [12] can help reduce VRS. The Sensory Conflict Theory assumes a mismatch between different sensory systems, which can cause sickness. For example, when there is visual movement (vection) but no vestibular stimulation (that is, seeing things and the environment moving but one is not physically moving), people can feel disoriented. One of the factors associated with its severity is the movement performed in the virtual world [13], [14]. Therefore, applications that require movement and generate vection seem to increase users' likelihood of feeling sick. As such, these theories support the use of one of the most well-known techniques for locomotion in VR (Teleport [2]), which does not require movement and is embedded in many commercial applications (e.g., Superhot VR [15]).

Other types of techniques that do not require reduced locomotion have also been proposed to reduce VR sickness [11], [16], [17]. Some involve altering some other visual aspects of the virtual environment and have been investigated to determine their relative performance and advantages [18]–[20]. These techniques can be grouped into two: (1) *visual* and (2) *locomotion*. Visual techniques do not explicitly alter movement, other physics properties, and environmental mechanics. Examples involve changing the viewing perspective or adding a gazing point. Locomotion techniques, on the other hand, alter in some way the physics to distort movement. Examples include Teleport and geometry deformation [21].

*A. Visual Techniques*

Visual techniques used to mitigate VRS vary widely in how they change the presentation of the VE. Some add information to the environment, such as Gaze Point [12], which gives users a central fixed rest frame. This technique is based on the Rest Frame Theory, which predicts that people get VRS because it is harder to find a stationary reference frame to rely on when navigating in VR. This technique takes different forms, such as a grid pattern [22], clouds in the sky [23], and minions [12].

Other visual techniques specific to HMDs are viewpoint snapping [24] and rotation blurring [25]. Both reduce the players' amount of virtual movement while keeping the physics laws aligned with the real world. To an extent, they are the equivalent of closing one's eyes while virtual movement is performed. Because no real movement happens and the virtual movement is not seen, there is no conflict between the senses. In their studies, they reduced the mean VRS by around 40%.

Many visual mitigation techniques are based on reducing users' field of view (FoV). FoV reduction has been widely studied and has led to positive results [4], [26]–[28]. It has been tested in simple tunnels [29] and can be inserted inside games. Also, it can sometimes be used in dynamic ways by, for example, proportionally changing the areas reduced with increased movement [27]. Yet, some people cannot adapt to FoV reduction [26], which can help reduce VRS by around 30%.

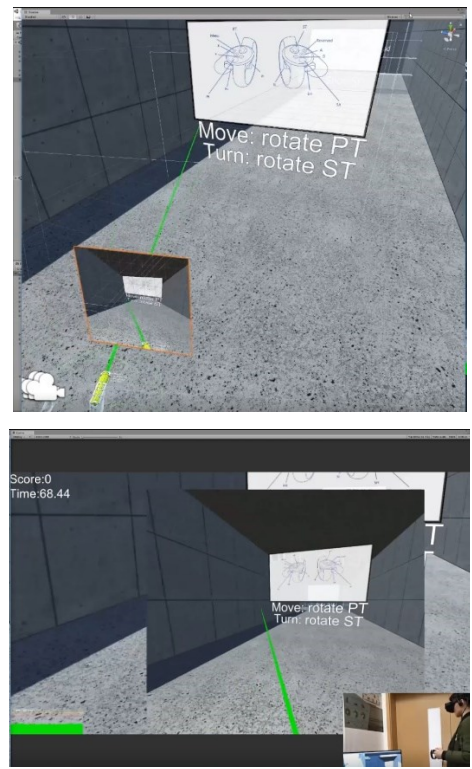

Figure 1: Upper part) The concept used in the 2D View technique as seen from the game editor; a 2D picture of the 3D environment is presented in front of the user's view. The 2D view is of configurable size. Lower part) The game implementation of the concept of the 2D view as seen from the player's eyes.

A recent paper [6] introduced a technique that combines the features of various visual techniques, including Gaze Point and FoV reduction. It provides a "picture" of the 3D virtual environment displayed as a 2D view (see Fig. 1) and has been shown advantages over a simple gaze point. Its positive results were due to the use of a 2D frame to view the 3D virtual environment. In some instances, the mean VRS reduction could be as high as 50% [6].

*B. Locomotion Techniques*

Geometry deformation [21] is a recent location technique. It changes the objects' mesh when the users walk towards them to reduce the virtual space that the users perceive as movement in the VR environment. However, because it is relatively new, it is unclear how well it would be accepted in a gaming context.

Teleport is arguably the most well-known locomotion technique. Teleport moves the player immediately from one place to another in a discrete way (see Fig. 2). It often uses a cut scene and removes the movement through the virtual world. Because no movement is present or seen by users, the level of sensory conflict is minimal, and hence the level of VRS is reduced. Some studies show that Teleport can be disadvantageous for performing VR tasks because it can be disorienting [30] and, as such, it may not be an appropriate choice for games. On the other hand, other researchers have not found any issues with disorientation and suggested that Teleport is more enjoyable in games when compared to other forms of locomotion [31]. Even though Teleport presented mixed results in user experience [30], [31], and some people even find it more sickening [32], [33], the mean reduction in VRS when using Teleport is around 50% [32], [34], which is similar to the best results of the 2DF [6].

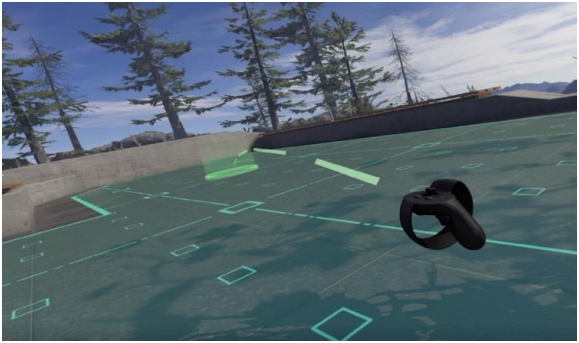

Figure 2: A screenshot of the standard Teleport technique used in the introduction of a commercial HMD.

The above findings led us to the following three research questions: (*RQ1*) *What are the trade-offs in immersion, VRS, and player performance when using Teleport in a fast-paced game*? (*RQ2*) *Is Teleport a more effective mitigation technique than 2DF*? (*RQ3*) *Can Teleport led to better performance than a 2DF like technique*?

### III. EXPERIMENT DESIGN

We conducted an experiment to evaluate how Teleport would behave in a first-person shooter (FPS) game and compared it to the 2D view reduction technique [6] in terms of VRS, user performance, and immersion. We chose the 2DF technique to compare against Teleport because it has performed better than other similar techniques [6].

We implemented the 2DF technique in Unity, as described in [6]. Teleport's implementation was based on the "Point & Teleport with Direction Specification" approach [35]. We used the same configurations of the standard SteamVR Teleport, transition style, and speed.

#### A. Game Environment

The game for our experiment was developed in-house to avoid or minimize confounding factors. It was a typical FPS game in which players were required to navigate rapidly and destroy the enemies but avoid being hit by them to finish the game. It consisted of a maze for players to traverse through. There were hallways, chambers, and pillars that can be used strategically within the maze. The adversaries were ellipsoids that shoot "energy balls" at the player. They would all attack upon locating the player's avatar.

The maze was composed of tall grey walls and, to reduce path memorization, it did not have other visual cues. Turns were designed to force the player to rotate during the game to increase the possibility of VRS [36]–[39]. Moreover, the pillars and chambers for hiding were meant to encourage the players to perform actions and balance shooting and hiding. There were no ambiguous paths or shortcuts. Consequently, all the players would follow roughly the same path.

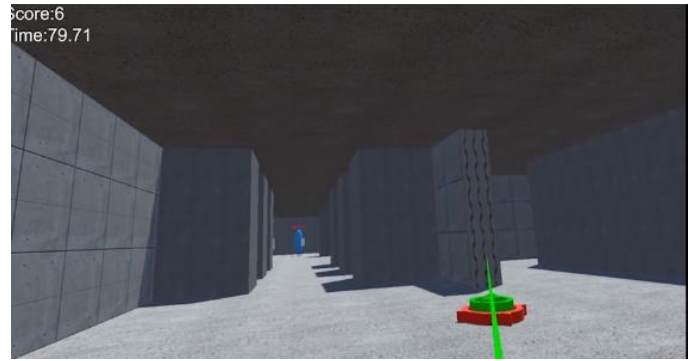

Figure 3: Our teleport method; a player can see the destination position (lower orange circle), and orientation (upper green circle). The enemies and walls follow the specifications given in [6] (i.e., monochromatic ellipsoids and grey walls).

In the first turns, we placed each kind of ellipsoid individually to give the players the chance to get familiar with their behavior and environment. After, the pace was increased by expanding the number of enemies up to the final hall, where most enemies were found (10).

#### B. Evaluation Metrics

The Simulator Sickness Questionnaire (SSQ) [40] and Immersive Experience Questionnaire (IEQ) [41] were used in the experiment. SSQ covers three parts to measure the level of *Nausea*, *Disorientation*, and *Oculomotor*. The overall weighted level for VRS (*Total Severity*) was calculated as the sum of these three parts. Each symptom uses a 5-point Likert Scale from least severe to most severe. IEQ is also implemented this way.

The following scoring system was used in the game: the player would get 10 points for hitting a target and lose 10 points if hit by an enemy. The score can be negative. Based on this system, Accuracy was defined as the number of shots that hit the target divided by the total shots.

We expected the players to be aware that the adversaries could kill them. However, as the experiment's goal was to investigate several facets of the gameplay experience, we set the players' health points (HP) to be able to endure at least 4 minutes of getting hit directly to guarantee the collection of enough meaningful data.

We also introduced an additional measurement: *Jumps*. Jumps counts the number of changes in the player's translational position after hitting a target, i.e., the number of different positions a player has when hitting different targets (see Fig. 4). We treated different positions between two adjacent shots as one Jump. We expected this measurement to inform us of the players' movement behavior, reflecting how careful and focused participants were during the game. Also, it can be an influential factor in terms of comparing continuous and discrete movements. We used the Jumps measurement to explore correlations with the other evaluated scales. Jumps were recorded in each experiment individually.

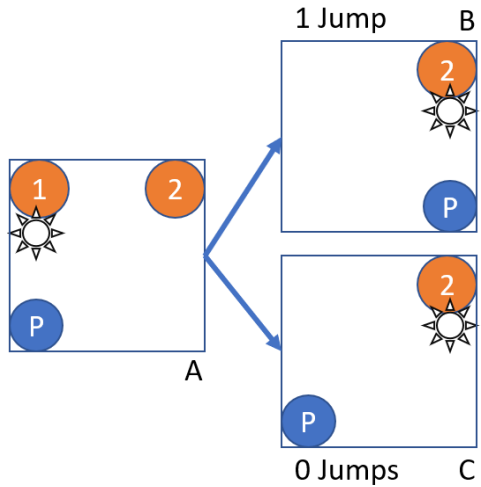

Figure 4: An illustration of the concept of Jumps. A) Initial state in which Player P is in its first registered position when Enemy 1 is hit; B) State in which one Jump is counted because, when the bullet hits Enemy 2, Player P's position is different from its original position in A); C) State in which no Jumps are counted because, when the bullet hits Enemy 2, Player P's position is the same its original position in A).

## IV. EXPERIMENT

Our experiment was designed to compare how Teleport and 2DF fare against each other and no VRS mitigation technique, which was our baseline technique. For Teleport, along with the instantaneous shift in position to the pointed area, the user's landing orientation after teleportation could be changed by adjusting the direction of the arrow in the expected position of the teleportation (see Fig. 3). We chose to use 2DF surrounded by a dark background because it was reported to be the most effective [6]. Hence, we evaluated the three conditions: 3D view and VR HMD with Teleport (Teleport), 2D frame in a VR HMD (2DVR), and 3D view and VR HMD without Teleport (3DVR).

In addition, as stated earlier, to compare these versions from a VRS, user performance, and subjective immersion perspective, we used *Jumps*. Since it is possibly a new measurement tool, we analyzed correlations between Jumps and the other measurements.

### A. Participants

We recruited a total of 18 participants (9 females) from a local university with an average age of 20.0 (s.d. = 1.90), ranging from 18 to 27. All volunteers had normal or corrected-to-normal vision. None of them declared any history of color blindness or health issues. Thirteen participants (72.2%) had experience with VR systems. Three (16.7%) reported that they felt sick when playing regular FPS games on a PC.

### B. Apparatus

We used an Oculus Rift CV1 as our HMD because it is one of the most popular off-the-shelf VR devices. A desktop with 16GB RAM, an Intel Core i7-7700k CPU @ 4.20GHz, a GeForce GTX 1080Ti dedicated GPU, and a standard 21.5" 4K monitor were used to drive the HMD. Also, we used a pair of Oculus Touch as our controller.

### C. Experimental Procedure

When participants reached the experimental room (an indoor lab on campus), they were asked to fill in a pre-experiment questionnaire to collect demographics and past gaming experience information. We assigned a specific order of these three versions for participants to follow using a Latin Square design to mitigate the accumulative sickness as much as possible. To let participants become familiar with the VR HMD and the Oculus Touch, we demonstrated the Oculus Rift using an official game demo without walking. We also introduced the rules of the game and the corresponding operations of the different versions.

Participants followed the defined order to play the three versions. After completing each version, they were asked to answer the questionnaires mentioned above and rest as much as they wanted before continuing the next version.

### D. Results

We used both statistical inference methods and visuals to analyze the data. A Shapiro-Wilk test was conducted first to check for the normality of the data. We used parametric tests for normally distributed data; for the others, we used non-parametric tests. We conducted Mauchly's Test of Sphericity for normally distributed data and employed Repeated Measures ANOVA (RM-ANOVA) with Bonferroni correction to detect overall significant differences. We applied the Greenhouse-Geisser correction when faced with a violation of the assumption of sphericity. For non-normal data, we conducted the Friedman test. If there was a significant difference, we then ran Wilcoxon signed-rank tests with Benjamini-Hochberg correction to pinpoint where the difference occurred among different combinations. Finally, we used Pearson's correlation for normally distributed data and Spearman's correlation for non-normal ones to identify correlations.

#### 1) Simulator Sickness Questionnaire (SSQ)

The results of the SSQ data are summarized in Fig. 5. The data were distributed normally in the four sub-scales. After conducting the Friedman Test, we found a statistically significant difference in all four sub-scales: $\chi^2(2) = 8.391$, $p = .015$ for Nausea, $\chi^2(2) = 6.578$, $p = .037$ for Oculomotor, $\chi^2(2) = 11.577$, $p = .003$ for Dizziness, and $\chi^2(2) = 6.491$, $p = .039$ for Total Severity (or simply Total).

Post-hoc test showed no significant difference among versions in Nausea ($Z = -1.848$, $p = .065$ for 3DVR and Teleport and $Z = -.205$, $p = .838$ for Teleport and 2DVR) except 2DVR and 3DVR, for which $Z = -2.672$, $p = .008$. Median (IQR) Nausea levels were 3, 2, and 5 for 2DVR, Teleport, and 3DVR, respectively.

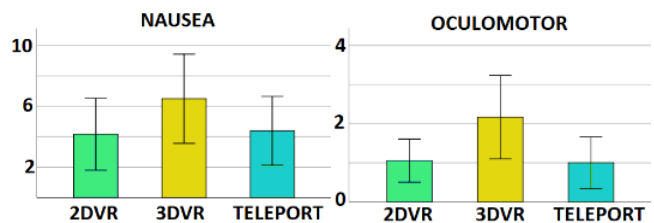

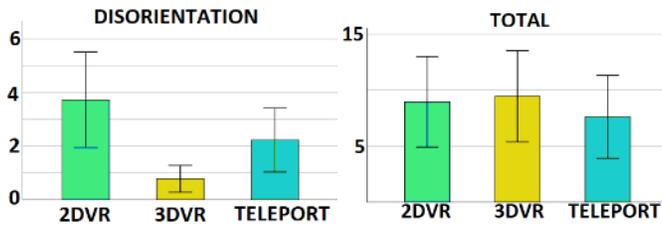

Figure 5: Mean results of SSQ concerning the four sub-scales of the three techniques. Bars represent 95% confidence intervals.

Next, we observed that in Oculomotor, there was a significant difference between 3DVR and Teleport (Z = -2.129, p = .033) and between 3DVR and 2DVR (Z = -2.183, p = .029) after correction. However, the difference between Teleport and 2DVR was not significant (Z = -.144, p = .885). Median (IQR) Oculomotor levels for 2DVR, Teleport, and 3DVR were 1, 0, and 2, respectively.

In terms of Disorientation, after we applied the same correction, they all had a significant difference with each other. In detail, Z = -2.186, p = .029 for 3DVR and Teleport, Z = -3.051, p = .002 for 3DVR and 2DVR, and Z = -2.214, p = .027 for Teleport and 2DVR. Their Median were 2DVR = 2.5, Teleport = 1, and 3DVR = 0.5, respectively.

Finally, there was no significant difference among different comparisons in Total: 3DVR and Teleport (Z = -1.122, p = .262), 3DVR and 2DVR (Z = -1.156, p = .248) and Teleport and 3DVR (Z = -1.226, p = .220). The Median IQR were 2DVR = 6, Teleport = 4, and 3DVR = 6.5.

*2) User Performance*

Fig. 6 shows the data for Score, Accuracy, Time, and Jumps of the three versions. For the score, Friedman Tests showed an overall significant difference among these versions ($\chi2(2)$ = 19.972, p =.000). Similarly, Jumps also differed significantly ($\chi2(2)$ = 19.681, p =.000). Similarly, RM-ANOVA tests revealed that Time differed statistically significantly (F (2, 34) = 4.260, p = .022, $\eta p^2$ = .200). However, a significant difference was not found for Accuracy ((F (2, 34) = 1.299, p = .286, $\eta p^2$ = .071).

In our post-hoc analysis, Wilcoxon signed-rank tests showed that Score in all versions differed significantly with each other: 3DVR and Teleport (Z = -3.528, p = .000), 3DVR and 2DVR (Z = -2.369, p = .018), and Teleport and 2DVR (Z = -3.158, p = .002). 74.5, 7.5, 60.5 were the Median (IQR) Score levels for 2DVR, Teleport and 3DVR, respectively.

Our post-hoc analysis also revealed that for Jumps there was a significant difference between 2DVR and Teleport (Z = -3.681, p = .000), and between Teleport and 3DVR (Z = -3.485, p = .000). However, no significant difference was found between 3DVR and 2DVR (Z = -1.308, p = .191). Their Median were 3DVR = 74.5, Teleport = 23 and 2DVR = 66.5.

In terms of Time, there was no significant difference between 3DVR and Teleport (p = .129), between 3DVR and 2DVR (p = 1.000), and between Teleport and 2DVR (p = .066).

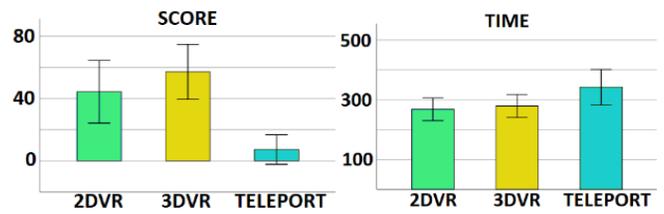

Figure 6: Mean results of Score, Jumps, Time, and Accuracy of the three techniques. Bars represent 95% confidence intervals.

*3) Perceived Immersion*

RM-ANOVA tests showed a significant difference in overall immersion (F (2, 34) = 5.520, p = .008, $\eta p 2$ = .245). Post-hoc analysis showed that immersion in 3DVR was significantly higher than in 2DVR (p = .017). However, there was no significant difference between 3DVR and Teleport (p = .187) or Teleport and 2DVR (p = .604). Fig. 7 shows the data for immersion among three versions.

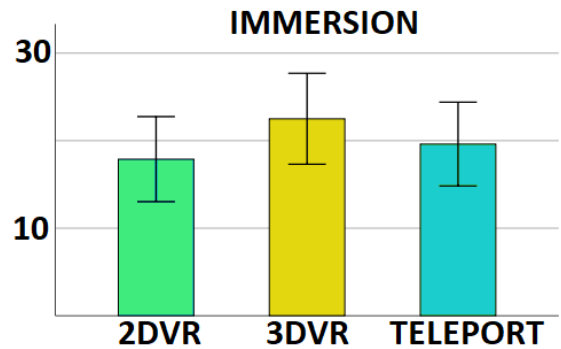

Figure 7: Mean immersion levels of the three techniques in the experiment. Bars represent 95% confidence intervals.

*4) Correlations between Jumps and other sub-scales*

Table 1 shows a summary of the correlations between Jumps and other sub-scales for the different versions. It was clear that in all three versions, there were strong and positive correlations between Jumps and Score, which were statistically significant: $r_s(8)$ = .824, p = .000 in 3DVR, r = .866, p = .000 in Teleport, and $r_s(8)$ = .847, p = .000 in 2DVR. It also showed that Jumps is not strongly and positively correlated with Time except in 2DVR ($r_s(8)$ = .621, p = .006). For Immersion, there was a strong and positive correlation in 3DVR and 2DVR, where $r_s(8)$ = .613, p = .007 and r = .485, p = .041, respectively.

For other sub-scales (Nausea, Oculomotor, Disorientation, Total and Accuracy), there were no significant correlations.

TABLE I: Summary of the results of the correlation coefficients, significant differences, types (S represents Spearman's correlation and P represents Pearson's correlation) and whether significant.

| Scale | r /rs | p | Type |
|---|---|---|---|
| Score-3DVR | .824 | .000 | S |
| Score-Teleport | .866 | .000 | P |
| Score-2DVR | .847 | .000 | S |
| Time-2DVR | .621 | .006 | P |
| Immersion-3DVR | .613 | .007 | S |
| Immersion-2DVR | .485 | .041 | P |

## V. Discussion

The results show differences in Nausea, Oculomotor, and Disorientation among the three techniques or versions. First, we did find 2DVR, the version based on the 2D frame technique, was better than 3DVR in Nausea and Oculomotor, indicating that 2DVR could make users feel lower VRS. This finding is aligned with the expectation that Teleport performed better than 3DVR in Oculomotor. 2DVR could help mitigate some VRS symptoms; however, it could cause more disorientation than the other versions. Because disorientation is supposedly one of Teleport's disadvantages [30], it is surprising that it was not the worst alternative. Based on the Total score, we did not find it either to be particularly superior in mitigating VRS.

Interestingly, even if Teleport did perform statistically better in some aspects of VRS, its Score and Jumps were significantly lower than the other versions. This result indicates that participants could not use Teleport well in a fast-paced environment, especially for FPS games requiring intense operations and a high concentration level. However, we failed to find statistical differences between Accuracy in Teleport and the other conditions. In other words, these different techniques did not stand out regarding their effect on aiming and shooting.

For immersion, 3DVR scored higher than 2DVR, which diverged from the results found in [6]. This difference could have resulted in the difference in controllers. We used a handheld controller, whereas the researchers in [6] used a keyboard and mouse, the combination which led to the best performance in their experiment. We were unable to find statistically significant differences between Teleport and 3DVR.

By our definition, Jumps were considered a measurement of how active users were during gameplay. Hence, it was not surprising that players had higher scores when they moved more frequently. VRS did not appear to correlate with Jumps, which is somewhat unexpected since movement can be associated with VRS [42].

Our results suggest that both Teleport and 2DVR could alleviate VRS to some extent but through different mechanisms. The total changes in shooting position did not influence VRS levels directly. It is also worth noting that players with more jumps had a higher immersion in 3DVR and 2DVR, indicating that 2DVR could provide enough immersion for more active players.

Regarding the three research questions, besides finding that Teleport trades a higher score for lower levels of Nausea and Oculomotor and higher Disorientation, we were unable to detect other trade-offs (*RQ1*). Teleport was not much superior to 2DVR in reducing VSRS (*RQ2*). They both present a similar profile in that they have similar trade-offs in VRS. However, participants had worse scores with Teleport (*RQ3*). Movements (Jumps) seem to be positively correlated with the score but did not have an influence on Accuracy. This result shows that for players in these three versions, the more active they were, the better performance they would have, even if players using Teleport had the lowest average Scores.

From the above results, two main recommendations could be extrapolated. First, they suggest that 2DF could be a useful yet simple and low-cost technique that can be adopted for FPS games because it can lower or minimize VRS while keeping the gameplay experience unaffected. Second, they show that Jumps, a measurement for movement and action, could be used as a possible marker for bottlenecks in the design of maps because lower Jumps seem to be associated with lower scores (i.e., lower performance).

### A. Limitations

The study might have been underpowered to detect differences in completion time. On the other hand, the number of participants is within the normal for this kind of experiment [43], and it had enough power to detect other important factors in the game experience.

In this study, we did not evaluate different frequencies of teleportation and variations of Teleport. However, this does not invalidate our study, given that we applied one of the most standard variations of Teleport. The number of teleportations performed by the players showed little variation, as presented in the Jump sections, which might indicate that they are within what can be expected for normal gameplay.

## VI. Conclusion

In this paper, we explored the effect of Teleport and a visual reduction technique on players' performance and gameplay when playing a first-person shooter (FPS) game in virtual reality (VR). In particular, we assessed whether they could help lower the level of VR sickness (VRS) and their trade-offs. Our results suggest that, although Teleport is suitable for mitigating VRS, it could also lead to lower performance compared to the other technique, which reduced the 3D view into a 2D view. Teleport led to lower scores and limited movement while only reducing VRS by just a small margin than the 2D technique. Our results show that the 2D reduction technique is useful and low-cost and can be adopted for FPS games because it can lower or minimize VRS symptoms while keeping the gameplay experience unaffected.

## Acknowledgments

The authors would like to thank the participants for their time and the reviewers for their reviews. The work is supported in part by Xi'an Jiaotong-Liverpool University (XJTLU) Key Special Fund (KSF-A-03; KSF-P-02) and XJTLU Research Development Fund.